\documentclass[aps,prb,twocolumn,nofootinbib,superscriptaddress,10pt]{revtex4-2}
\usepackage{libertine}
\usepackage[T1]{fontenc}
\usepackage[libertine,upint]{newtxmath}
\usepackage[cal=stixtwofancy,frak=stixtwo]{mathalfa}
\usepackage{graphicx,microtype,mathtools,bm,floatrow,xcolor,cancel}

\usepackage{ragged2e}

\makeatletter
\long\def\@makecaption#1#2{%
  \vskip\abovecaptionskip
  \sbox\@tempboxa{#1. #2}%
  \ifdim \wd\@tempboxa >\hsize
    \begin{minipage}{\hsize}\justifying
      #1. #2\par
    \end{minipage}%
  \else
    \centerline{#1. #2}%
  \fi
  \vskip\belowcaptionskip
}
\makeatother

\usepackage[caption=false,font=footnotesize]{subfig}

\definecolor{linkscolor}{cmyk}{0.6, 0.3, 0, 0.9}
\usepackage[colorlinks=true,allcolors=linkscolor!60]{hyperref}

\DeclarePairedDelimiterX\braket[2]{\langle}{\rangle}{#1 \delimsize\vert #2}
\DeclarePairedDelimiterX\matrixel[3]{\langle}{\rangle}{#1 \delimsize\vert #2 \delimsize\vert #3}
\DeclarePairedDelimiter\abs{\lvert}{\rvert}

\DeclareMathOperator{\sgn}{sgn}
\DeclareMathOperator{\const}{const}

\DeclareMathOperator{\im}{Im}

\NewDocumentCommand{\grad}{e{_^}}{%
    \mathop{}\!% \mathop for thin spacing before \nabla
    \nabla
    \IfValueT{#1}{_{\!#1}}% tuck in the subscript
    \IfValueT{#2}{^{#2}}% possible superscript
}

\begin{document}
\captionsetup[figure]{justification=justified,singlelinecheck=false}

\title{Collective excitations and stability of a non-Fermi liquid state near a quantum-critical point of a metal}

\author{Yasha Gindikin}
\affiliation{W.I.\ Fine Theoretical Physics Institute and School of Physics and Astronomy, University of Minnesota, Minneapolis, Minnesota 55455, USA}
\author{Dmitrii L.\ Maslov}
\affiliation{Department of Physics, University of Florida, Gainesville, Florida 32611--8440, USA}
\author{Andrey V.\ Chubukov}
\affiliation{W.I.\ Fine Theoretical Physics Institute and School of Physics and Astronomy, University of Minnesota, Minneapolis, Minnesota 55455, USA}

\begin{abstract}
We examine the spectral properties of collective excitations with finite angular momentum $l$ for a system of interacting fermions near a Pomeranchuk quantum critical point, both in the Fermi liquid and non-Fermi liquid regimes. Previous studies found that deep in the Fermi liquid regime, the spectral functions for even and odd $l$ behave differently---the latter is suppressed compared to the former because of kinematic constraints on scattering processes. The main focus of our paper is to understand how the spectral functions for even and odd $l$ evolve as the system enters the non-Fermi liquid regime. We obtain the full scaling function for the electron polarization bubble at arbitrary $l$, which interpolates between the Fermi liquid and non-Fermi liquid regimes. We show that collective excitations for all $l$  remain stable and causal throughout the crossover and right at the quantum critical point.
\end{abstract}

\maketitle
\textit{Introduction.}~
Understanding a non-Fermi liquid (NFL) state of an interacting electron system remains one of the most pressing challenges of modern condensed matter physics (see~\cite{PALee1989,Altshuler1994,Metzner2003,Abanov2003,PhysRevB.82.075127,hartnoll_2018,Lee2018,Chowdhury_22} and references therein). One of the key issues here is whether a NFL state remains stable over a wide energy range, even if the interactions inevitably drive the system toward superconductivity at sufficiently low energies.

A paradigmatic route to realizing a NFL is to tune the system towards a quantum critical point (QCP), associated with charge or spin ordering. Close to a QCP, electrons couple strongly to soft fluctuations of the corresponding order parameter. In $D \leq 3$, exchange by soft bosonic fluctuation between fermions leads to a singular self-energy that destroys quasiparticle coherence. One of the most studied examples is an Ising-nematic QCP~\cite{PhysRevB.50.17917,Oganesyan2001,Metzner2003,DellAnna2006,Rech_2006,Maslov2010,PhysRevB.82.075127,Lederer2017,Lee2018,Zhang_23,*Raines_24,Nosov2023,PhysRevB.89.155130,*PhysRevB.103.235129,*PhysRevB.106.115151}, which breaks rotational symmetry but leaves translational symmetry intact~\cite{pomeranchuk1958stability}. In a two-dimensional (2D) system, exchange by critical fluctuations of the nematic order parameter yields the electron self-energy $\Sigma^{'} (\omega) \sim \Sigma^{''} (\omega) \propto \omega^{2/3}$. At small $\omega$, $\Sigma(\omega)$ exceeds the bare $\omega$ in the fermionic propagator, which is a signature of a NFL behavior.

Existing theories of a NFL behavior near a QCP assume that a NFL state is stable with respect to perturbations~\footnote{A pairing instability generally does develop, but it has been argued~\cite{Raghu_2010,Fitzpatrick_2014,Lee2018,Efetov_2013} that at least in some cases it can be either  eliminated or moved to a very low energy by a proper choice of parameters}. The subject of this communication is a verification of this assumption. A given state of interacting fermions is stable if all perturbations decay with time. This holds if the excitations are causal, that is, if the corresponding dynamical susceptibilities are analytic in the upper half-plane of the complex frequency~\cite{LandauLifshitz5}.

In our study, we consider a system with an isotropic fermionic dispersion, when particle-hole excitations are split into decoupled channels with different angular momenta $l$. We assume that the system is close to a Pomeranchuk instability in a channel $l = l_0$ and consider the damping rates of particle-hole excitations in channels with $l \neq l_0$ due to strong interaction between low energy fermions and near-critical excitations in the $l=l_0$ channel. To avoid unnecessary complications associated with the angular dependencies of this interaction and of the dynamical part of the propagator of a near-critical boson, we assume that the system is close to a charge instability at $l_0 =0$ (the one that leads to phase separation~\cite{Mayrhofer_24}) and analyze particle-hole susceptibilities $\chi_l (q, \omega)$  in channels with  $l >0$. We conjecture that the results that we obtain near the $l=0$ instability hold for the cases of other $l_0$, including the most studied case of a system near an Ising-nematic instability ($l_0 =2$) (see e.g., Refs.~\cite{Oganesyan2001,Metzner2003,DellAnna2006,Rech_2006,Maslov2010,Lee2018,Zhang_23,*Raines_24,PhysRevB.89.155130,*PhysRevB.103.235129,*PhysRevB.106.115151}). Because susceptibilities at $l >0$ are not protected by conservation laws, the spectral properties of $\chi_l(q,\omega)$  with a generic $q$ are qualitatively the same as at $q\to 0$. We focus on this limit because the susceptibility $\chi_l(q \to 0,\omega)  \equiv \chi_l(\omega)$ is related to the dynamical part of the polarization bubble, $\Pi_l (\omega)$, with the same $l$:
\begin{equation}
\label{1_c}
  \chi_l(\omega) \propto \frac{1}{1 - a_l \Pi_l (\omega)}\,,
\end{equation}
where $a_l >0$~\footnote{This relation is consistent with a random-phase-approximation (RPA) formula $\chi_l(\omega) = \Pi^{\rm{tot}}_l(\omega)/(1 + U_l \Pi^{\rm{tot}}_l(\omega))$ once we split the total bubble $\Pi_l^{\rm{tot}}(\omega)$ into the static and dynamic parts as $\Pi_l^{\rm{tot}}(\omega)=\Pi_{l} (0) + \Pi_l (\omega)$, and expand in $\Pi_l (\omega)/\Pi_l (0)$. This yields $a_l = \left[\Pi_l (0) (1 + U \Pi_l (0))\right]^{-1}$. At the RPA level, excitations in $l \neq 0$ channels are stable when $\Pi_l (0) >0$ and $ 1 + U \Pi_l (0) >0$. Under these conditions, $a_l >0$.}.
For a nonzero $q$, this does not hold as $\chi_l (q, \omega)$ is expressed via $\Pi_{l'} (\omega)$ with both $l'=l$ and $l' \neq l$, Ref.~\cite{PhysRevResearch.1.033134}. In what follows, we will compute $\Pi_l (\omega)$ near a QCP analytically and analyze the properties of $\chi_l (\omega)$ using Eq.~\eqref{1_c}. Causality requires that $\chi_l (z)$, viewed as a function of a complex frequency $z = \omega' + i \omega''$, is an analytic function of $z$ in the upper half-plane~\cite{LandauLifshitz5}. A necessary, but not a sufficient, condition for causality is the positivity of the dissipative part of the susceptibility, which, according to Eq.~\eqref{1_c} is the condition $\sgn \left[\Pi''_l (\omega)\right] = \sgn \omega$.

The form of $\Pi^{''}_l (\omega)$ in the FL regime both away and near a QCP has been analyzed in a series of recent studies~\cite{Levitov2019,Klein_2018,*Klein_2018_1,10.21468/SciPostPhys.13.5.102,
PhysRevB.89.155130,*PhysRevB.103.235129,*PhysRevB.106.115151,Li:2023,*PhysRevB.109.115156,*PhysRevB.110.085139,Ips,Guo}. Near a QCP, a FL behavior holds at frequencies $\abs{\omega} < \xi^{-3}$, where $\xi$ is the correlation length in the $l_0$ channel. Remarkably, the functional form of $\Pi''_l (\omega)$ is different for even and odd $l$. For even $l$, $\Pi''_l (\omega)$ scales as $\omega \xi^{2}$, whereas for odd $l$, the linear term is absent, and $\Pi''_l (\omega)$ scales as $\omega^3 \xi^4 \ln{(\abs{\omega}\xi^3)}$. This distinction arises from a kinematic constraint on scattering processes in odd-$l$ channels. The same constraint also leads to a suppression of the optical conductivity compared to the canonical FL form $\sigma(\omega)=\const$. Instead, $\sigma (\omega) \propto \omega^2 \xi^4 \ln (\abs{\omega} \xi^3)$~\cite{Rosch:2005,Rosch:2006,Sharma:2021,Li:2023,*PhysRevB.109.115156,*PhysRevB.110.085139}. Despite these differences, collective excitations remain causal in all channels; perturbations with both even and odd $l$ decay with time, i.e., a FL ground state is stable.

We extend the description of collective excitations into the NFL regime near a QCP, where $\abs{\omega} > \xi^{-3}$. At a qualitative level, one might be tempted to describe a crossover from the FL to the NFL regime by replacing $\xi$ in the FL results by ${(-i \omega)}^{-1/3}$ for $\omega >0$. (On the Matsubara axis, $\xi$ gets replaced by $\abs{\omega_m}^{-1/3}$.) Doing so, one would obtain $\Pi''_l (\omega) \sim \im (e^{i \pi/6}) \omega^{1/3}$ for even $l$ and $\Pi''_l (\omega) \sim \im (e^{-i \pi/6}) \omega^{5/3}$ for odd $l$. For even $l$, the prefactor is positive, and we argue that this procedure reproduces the actual result (see also Ref.~\cite{Klein_2018,*Klein_2018_1}). However, for odd $l$, $\Pi^{''}_l (\omega)$ is negative, i.e., describing the crossover this way, one would find that causality is violated in the NFL regime, and thus a NFL state is unstable toward some unidentified order.

We claim that the transformation from the FL to the NFL regime cannot be captured by a naive substitution of $\xi$ by ${(-i \omega)}^{-1/3}$ in the FL formula for $\Pi''_l (\omega)$. Instead,
one has to calculate the full scaling function for $\Pi''_l (\omega)$, which interpolates between the FL regime at $\omega \xi^3 <1$ and  the NFL regime at $\omega \xi^3 >1$ (we restore the dimensional factor below). We show that $\chi_l (\omega)$ with both even and odd $l$ remains causal at $\omega \xi^3  >1$, including right at the QCP, and thus the system is stable towards perturbations with any angular momentum $l$ both in the FL and NFL regimes. This conclusion differs from the one in Ref.~\cite{Guo}. We will point out the reason for the difference later on in the paper.

While our calculations are technically involved, the key message is transparent. In the FL regime, $\Pi_l(\omega)$ admits a series expansion  in powers of $\omega\xi^3<1$. The leading term is $\Pi^{'}_l (\omega) \propto \omega^2 \xi$, while the subleading term for odd $l$ channels is $\Pi^{''}_l (\omega) \propto \omega^3 \xi^4$. As the system crosses over from the FL to the NFL regime, contributions of higher orders in $\omega \xi^3$ become relevant. These contributions determine $\Pi_l (\omega)$ as a scaling function of $\omega \xi^3$. We compute the full scaling function exactly. At a QCP, it yields $\Pi_l (\omega) \propto e^{i(\pi/6) \sgn \omega} \abs{\omega}^{5/3}$. We see that the sign of  $\Pi^{''}_l (\omega)$ at the QCP coincides with the sign of $\omega$, as required by causality. We show that the same holds in the whole crossover region between the FL and the NFL\@. We also show that at any distance from the QCP and even right at the QCP, $\chi_l (\omega)$ is analytic in the upper half-plane of complex frequency, i.e., $\chi_l (t<0) =0$.

\textit{Model.}~
We consider a rotationally-invariant 2D fermionic system near a $l=0$ Pomeranchuk instability in the charge sector. The precise form of the isotropic fermionic dispersion does not matter for our purposes,
and we just chose it parabolic.

The main object of our study is the dynamical part of the fully dressed particle-hole bubble at zero incoming momentum, $\Pi_l (\omega)$, with external vertices $\mathcal{V}_{\!l} (\bm{k})$. We approximate the external vertices by their values for $\bm{k}$ on the Fermi surface and choose $\mathcal{V}_{\!l} (\bm{k}) = \cos(l \theta_{\bm{k} \hat{x}})$, where $\theta_{\bm{k} \hat{x}}$ is the angle between $\bm{k}$ and a vanishingly small $\bm{q}$, which for definiteness we direct along $x$~\footnote{Note that for $l=1$, our polarization bubble differs from the current-current correlator $K_1$, for which  $\mathcal{V}_{\!l=1} (\bm{k}) = \bm{k}$. Whereas $K_1\vert_{q=0,\omega\neq 0}=0$ by current conservation~\cite{PhysRevB.50.17917,Maslov_2017}, our $\Pi_1 (\omega)$ is finite at $q=0$. We have checked that $\Pi_1 (\omega) =0$, if the side vertices are replaced by $\mathcal{V}_{\!1}(\bm{k}) = \bm{k}$.}.

\begin{widetext}

\begin{figure}[htb]
  \centering
  \subfloat[]{\includegraphics[width=0.19\textwidth]{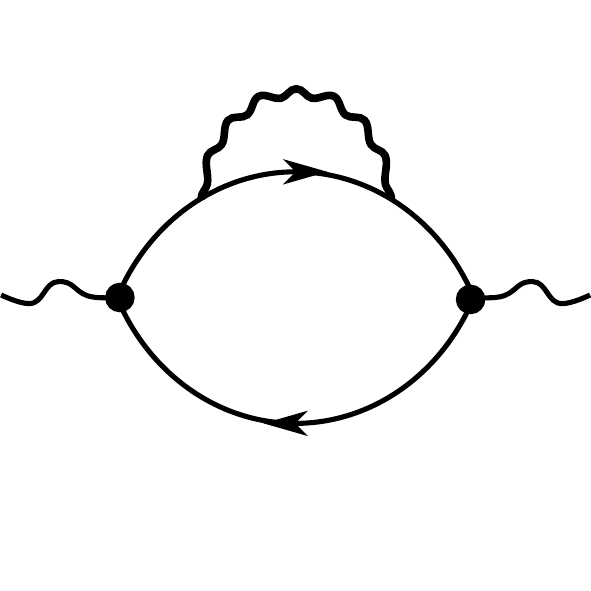}\label{fig:se1}}\hfill
  \subfloat[]{\includegraphics[width=0.19\textwidth]{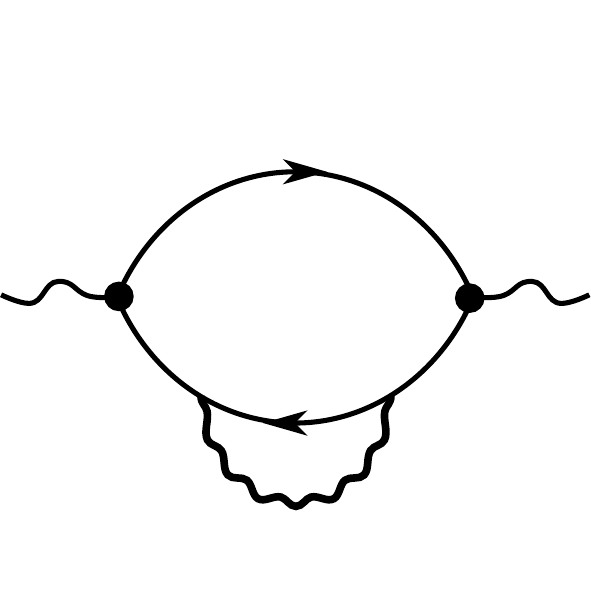}\label{fig:se2}}\hfill
  \subfloat[]{\includegraphics[width=0.19\textwidth]{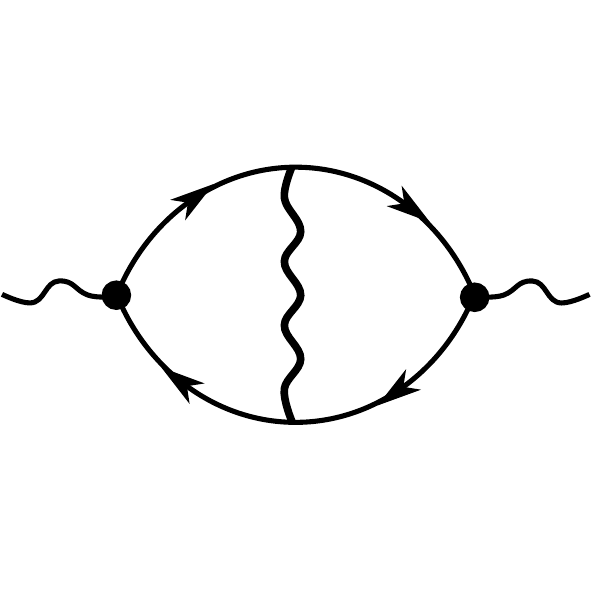}\label{fig:MT}}\hfill
  \subfloat[]{\includegraphics[width=0.19\textwidth]{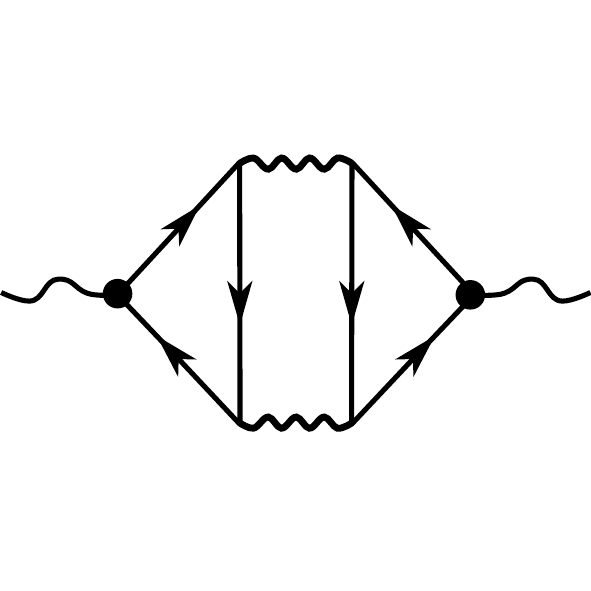}\label{fig:AL1}}\hfill
  \subfloat[]{\includegraphics[width=0.19\textwidth]{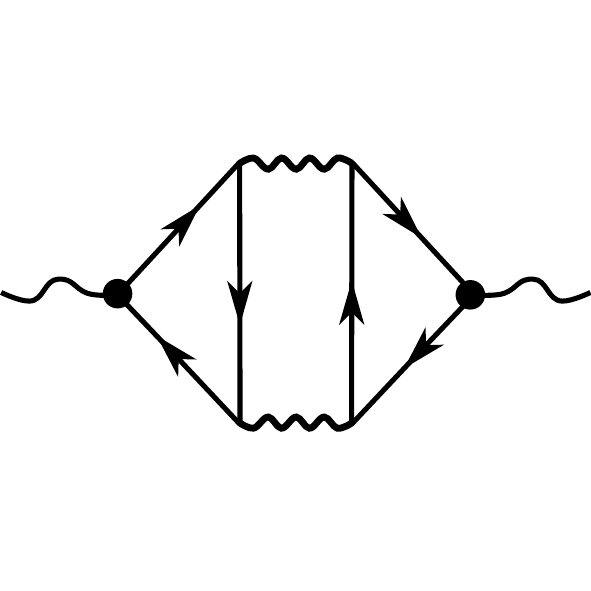}\label{fig:AL2}}
  \caption{%
    Diagrams for the polarization bubble~$\Pi_l$. In commonly adopted notations, diagrams (a,b) are the two self‐energy diagrams, diagram (c) is the Maki–Thompson diagram, and diagrams (d,e) are the two Aslamazov–Larkin diagrams. In all diagrams the external momentum $\bm{q}=0$, the external frequency $\omega$ is finite, and the side vertices (black dots) are $\mathcal{V}_{\!l}(\bm{k})=\cos(l\,\theta_{\bm{k}\hat{x}})$, where $\theta_{\bm{k}\hat{x}}$ is the angle between the internal fermion momentum~$\bm{k}$ and the $x$‐axis. The wavy line is the propagator of near-critical $l=0$ charge fluctuations, Eq.~\eqref{chi}. \label{fig:NC}}
\end{figure}

\end{widetext}

For free fermions, $\Pi_l (\omega) =0$ for all $l$ because, without interaction, a partial fermionic density along any direction of $\bm{k}$ is conserved~\cite{Klein_2018,*Klein_2018_1}. This approximate conservation law no longer holds in an interacting system: while $\Pi_{l=0} (\omega)$ remains zero due to the particle number conservation, $\Pi_{l\neq 0}(\omega)$ generally becomes finite.

In what follows, we present our results for $\Pi_l (\omega)$, obtained by collecting the contributions from an effective  4-fermion interaction mediated by a propagator of nearly-gapless bosonic fluctuations in the $l=0$ channel. We calculate $\Pi_l$ first on the Matsubara axis ($\omega \to \omega_m$), both analytically and numerically, and then rotate the Matsubara result to the real axis using $i\omega_m \to \omega + i\delta$. On the Matsubara axis, the propagator of nearly gapless $l=0$ bosonic fluctuations is given by
\begin{equation}
  \label{chi}
  \chi (p)  = \frac{\chi_0}{\bm{p}^2 + \xi^{-2} +\gamma \frac{v_F \abs{p_0}}{\sqrt{v_F^2\bm{p}^2+p_0^2}}}\,,
\end{equation}
where $p=(\bm{p},p_0)$, $p_0$ is the bosonic Matsubara frequency, $v_F$ is the Fermi velocity, $\gamma =\nu \bar{g}/v_F$ is the strength of Landau damping, $\nu=m/(2\pi)$ is the density of states at the Fermi energy, and $\bar{g}$ is the effective fermion-boson coupling with the dimensions of energy.

\textit{Particle-hole polarization bubble $\Pi_l (\omega)$.}~
 We obtain $\Pi_l (\omega)$ in two steps. First, we compute it with the minimum number of bosonic propagators that contribute to the same order in $\bar g$, while treating the internal fermions as free particles and neglecting vertex corrections. Then we show that insertions of the fermionic self-energies and vertex corrections cancel each other, at least if the self-energy can be approximated as being purely dynamic, and hence do not modify $\Pi_l (\omega)$ obtained at the first step.

To lowest order in the fermion-boson coupling, $\Pi_l (\omega)$ is represented by five diagrams~\cite{Maslov_2017}, see Fig.~\ref{fig:NC}: two diagrams with self-energy (SE) insertions into the fermionic lines,
Figs.~\ref{fig:se1} and~\ref{fig:se2},a Maki-Thompson (MT) vertex correction, Fig.~\ref{fig:MT}, and two  Aslamazov-Larkin (AL) diagrams, Figs.~\ref{fig:AL1} and~\ref{fig:AL2}.

Although the SE and MT diagrams each contain a single interaction line, while the AL diagrams contain two, all contributions to $\Pi_l (\omega)$ turn out to be of the same order. This is so because the contributions from
the SE and MT diagrams are nonzero only due to the Landau-damping term in the bosonic propagator $\chi(p)$, which itself originates from dressing the static bosonic propagator by particle-hole bubbles and includes the
fermion coupling $\bar{g}$ in its prefactor.

We evaluated all five diagrams. The sum of the SE and MT contributions is independent of whether the angular momentum $l$ is even or odd. In contrast, the sum of the two AL diagrams depends strongly on the parity of $l$.
The computations, though lengthy, are straightforward. We present the details in Appendix~\ref{Sec:appendix} and summarize the results here.

In the FL regime, the dynamical term in the bosonic propagator $\chi(p)$ can be expanded in powers of $\omega$. Doing so, we find
\begin{equation}
\label{Eq:MT+SE_re_1}
	\Pi_{l, MT+SE} (\omega)= i c_1 \frac{\bar{g}^2 l^2 \xi^2}{v^4_F} \omega
\end{equation}
and
\begin{equation}
\label{Eq:AL_re_1}
	\Pi_{l, AL} (\omega)= -\frac{c_1}{2} \left(1 - {(-1)}^l\right)
\frac{\bar{g}^2 l^2 \xi^2}{v^4_F} i\omega\,.
\end{equation}
Here and in what follows, $c_{1\dots 5}>0$ are numerical coefficients.

For even $l$, the AL contribution in~\eqref{Eq:AL_re_1} vanishes and $\Pi_l (\omega) \approx \Pi_{l, MT+SE} (\omega)$. This contribution is imaginary and linear in $\omega$. This result is in full agreement with Ref.~\cite{Guo}, including the numerical prefactor. Substituting $\Pi_l (\omega) \approx \Pi_{l, MT+SE} (\omega)$ into Eq.~\eqref{1_c} for $\chi_l(\omega)$ and extending $\omega$ to the entire complex plane $z = \omega' + i \omega''$, we find that $\chi_l (z)$ is analytic in the upper half-plane, hence $\chi_l(t <0) =0$.

For odd $l$, the situation is different, as $\Pi_{l, MT+SE} (\omega)$ and $\Pi_{l, AL} (\omega)$ cancel each other. A nonzero $\Pi_l (\omega)$ then comes from a subleading term. We find (see Appendix~\ref{Sec:appendix}) that this term comes entirely from the AL diagrams and gives
\begin{equation}
  \label{nn_1}
    \Pi_{l} (\omega) =  c_2  \frac{\bar{g}l^4  }{m v^5_F}  \omega^2 \xi
     + i c_3 \frac{ \bar{g}^2 l^4 }{v^6_F}
     \omega^3 \xi^4 \ln \frac{1}{
     \gamma \xi^3 \omega}\,.
\end{equation}

Next, we present the results in the opposite limit, when the system is at QCP\@. In this limit, one must keep the entire dynamical part of $\chi(p)$. For even $l$, we find that the leading contribution to $\Pi''_{l,QCP} (\omega)$ is again given by the sum of the SE and MT diagrams, while the two AL diagrams cancel each other. We find
\begin{equation}
        \Pi''_{l, QCP} (\omega)=
  c_4 \left(\frac{\bar{g}^{4} l^6}{m^2 v^{10}_F}\right)^{1/3} \abs{\omega}^{\frac13} \sgn{\omega}\,.
\end{equation}
This result matches the form of Eq.~\eqref{Eq:MT+SE_re_1} under the substitution $\xi\to\abs{\omega}^{-1/3}$. One can readily verify that the corresponding response function $\chi_{l,\, \text{QCP}}(\omega)$ is causal.

For odd $l$, the leading contributions from the SE+MT and AL diagrams again cancel each other, and the polarization bubble is given by a subleading term. Like in the FL regime, this term comes solely from the AL diagrams. We find
\begin{equation}
  \label{nn_5}
        \Pi''_{l, QCP} (\omega) = c_5 \frac{\bar{g}^2l^4}{v^6_F \gamma^{4/3}} \abs{\omega}^{\frac53}\sgn{\omega}\,.
\end{equation}

We see that $\sgn \left[\Pi''_{l,QCP} (\omega)\right] = \sgn \omega$, consistent with causality. Furthermore, adding the real part $\Pi'_{l, QCP} (\omega)$, substituting the full $\Pi_{l, QCP} (\omega)$ into $\chi_{l,QCP} (\omega)$, and extending $\omega$ to the entire complex plane of $z = \omega' + i \omega^{''}$, we find that $\chi_{l,QCP} (z)$ is analytic in the upper half-plane of $z$.

\begin{figure}
	\includegraphics[width=0.9\linewidth]{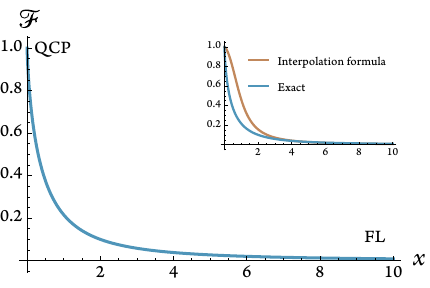}
	\caption{The scaling function $\mathcal{F}(x)$ from Eq.~\eqref{nn_6}, where $x = \omega_{FL}/\omega$. The limit $x=0$ corresponds to a QCP and $x \gg 1$ to the FL regime. The scaling function is normalized to $\mathcal{F}(0) =1$. In the inset we compare the exact $\mathcal{F}(x)$ and its approximate form presented in the text.
 \label{Fig:text}}
\end{figure}

With an extra computational effort, we obtained the full crossover scaling function for $\Pi''_l (\omega)$ for odd $l$, which interpolates between the FL result in Eq.~\eqref{nn_1} and the QCP form in Eq.~\eqref{nn_5}. The scaling function is defined as
\begin{equation}
  \label{nn_6}
  \Pi''_{l} (\omega) =  \Pi''_{l, QCP} (\omega) \mathcal{F}\left[{\left(\frac{\omega_{\mathrm{FL}}}{\omega}\right)}^{2/3} \right]
\end{equation}
where $\omega_{FL} = 2 \xi^{-3}/\gamma$. A FL behavior holds when $\omega \ll \omega_{FL}$ and a NFL behavior holds when $\omega \gg \omega_{FL}$. The function $\mathcal{F}(x)$ in shown in Fig.~\ref{Fig:text}. At $x \ll 1$, $\mathcal{F}(x) \approx 1 + 1.06 x \ln x$, at $x\gg 1$, $\mathcal{F}(x) \propto \ln{x}/x^2$. We present the derivation of the full scaling function in Appendix~\ref{Sec:appendix}. We see from the plot that the sign of $\Pi''_{l} (\omega)$ does not change in the entire ranges of frequencies. This implies that causality is preserved in the whole crossover range between the FL and NFL regimes. To a reasonably good numerical accuracy, $\mathcal{F} (x)$ can be represented by an extrapolation formula $\mathcal{F} (x) \approx 2 \im {\left(\frac{i}{1+ i x^{3/2}}\right)}^{1/3}$. We compare the exact and the approximate $\mathcal{F}(x)$ in the insert in Fig.~\ref{Fig:text}.

\textit{Effects of quasiparticle residue and vertex corrections.}~
We now demonstrate that the results obtained with free-fermion propagators and bare vertices do not change when we include the dynamical self-energy and vertex corrections because the two effects cancel each other. For definiteness, we consider the FL regime. Earlier studies have found (see e,g.,~\cite{Raines_24} and references therein) that an Eliashberg-type description is justified  near a QCP, which implies that the leading term in the self-energy at small $\omega$ is purely dynamical: $\Sigma = \lambda \omega$, where $\lambda \sim {\bar g}/(v_F \xi^{-1})$. The dressed fermionic propagator is $G(k,\omega) = Z/(\omega - Z \varepsilon_k)$, where $Z = 1/(1+\lambda)$ is the quasiparticle residue and $\varepsilon_k$ is the fermionic dispersion. The Eliashberg description is valid when $\xi$ is sufficiently large, such that  $\lambda \gg 1$ and $Z \ll 1$. For a proper description, $Z$ then must be kept in the fermionic propagator. We emphasize that these  renormalizations are different from the one in $\Pi_{SE} (\omega)$ in Figs.~\ref{fig:se1},\ref{fig:se2}. For the latter, momentum dependence of the self-energy, while subleading, is crucial, as without it one would not obtain a frequency-dependent part of $\Pi_{SE} (\omega)$. Furthermore, renormalization of the fermion-boson vertices is small within the Eliashberg theory, but renormalization of side vertices in all diagrams in Fig.~\ref{fig:NC} is not small and must be included.

\begin{figure}[htb]
    \centering
    \includegraphics[width=.4\linewidth]{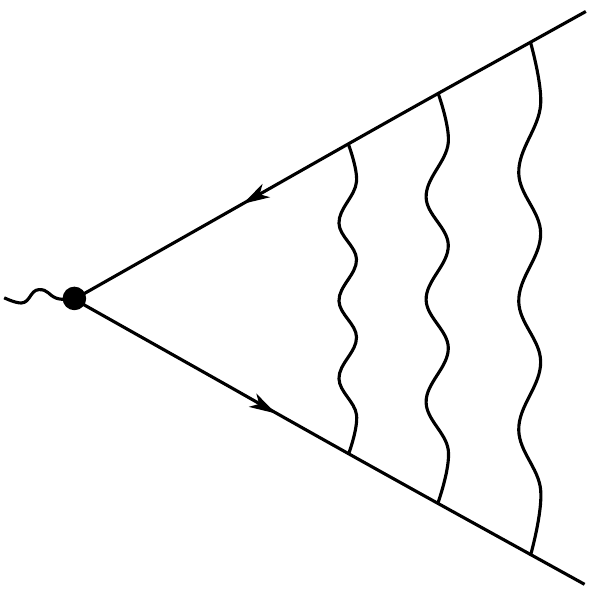}
    \caption{Ladder of vertex corrections for the side vertex in the particle-hole polarization. Each wavy line is the near-critical bosonic propagator. In the FL regime, it can be approximated as static. \label{fig:ladder}}
\end{figure}

To clarify this issue, we recall that in the $l=0$ channel, the Ward identity, associated with particle number conservation, equates the side vertex $\Gamma$ to $1 + d \Sigma/d\omega = 1+\lambda =1/Z$ (see, e.g.,~\cite{Mayrhofer_24} and references therein). In the Eliashberg theory, the Ward identity can be verified by summing up the ladder series of vertex corrections shown in Fig.~\ref{fig:ladder}. A single rung of the ladder (a convolution of two fermionic propagator and the interaction line) is given by $\lambda Z = \lambda/(1+\lambda)$. Summing up the geometric series, one obtains $\Gamma= 1/\left(1- \lambda/(1+\lambda)\right) =
1+\lambda$, as expected.

For our case of $l\neq 0$, the side vertices contain angular-dependent form-factors $\mathcal{V}_{\!l} (\bm{k})$, and there is no Ward identity as the corresponding modulated densities are not conserved. However, the momentum transfer in each scattering of a near-critical boson is small, of order $\xi^{-1}$. To leading approximation in $1/k_F\xi$,  the angular-dependent $\mathcal{V}_{\!l} (\bm{k})$ can be moved through the ladder series as if it were a constant. Without the $\mathcal{V}_{\!l} (\bm{k})$ term, the renormalization is the same as for $l=0$, i.e., the dressed vertex $\Gamma_l$ still scales as $1/Z$. We analyze this case more carefully in Appendix~\ref{Sec:appendix}.

We next discuss possible renormalization of the Landau-damping term in the near-critical bosonic propagator, Eq.~\eqref{chi}. For free fermions, the Landau-damping term on the Matsubara axis is given by $\Pi^{(0)}_0(p)  = \nu \abs{p_0}/\sqrt{v_F^2 \bm{p}^2 + p_0^2}$. To re-analyze this term for dressed fermions, we note that for free fermions this term comes from the polarization bubble made of fermions with momenta $\bm{k}-\bm{p}/2$ and $\bm{k}+\bm{p}/2$, once we integrate over the angle $\theta$ between internal $\bm{k}$ and external $\bm{p}$, i.e., from $ \Pi^{(0)}_0 (p) = \nu \int d\theta (2\pi) (p_0/(p_0 +i v_F \abs{\bm{p}} \cos \theta))$. We now show that to properly include the effect of vertex corrections to the bubble, one has to sum up ladder series of vertex corrections \textit{before} integrating over $\theta$. One can easily make sure that in the FL regime the ladder series for the vertex is geometric and sums to $\Gamma (p) = 1 + u + u^2 + \ldots=1/(1-u)$, where $u = \lambda p_0/(p_0/Z +i v_F \abs{\bm{p}} \cos{\theta})$. This gives
\begin{equation}
    \Gamma (p) = (1 + \lambda) \frac{p_0 + i Z v_F\abs{\bm{p}} \cos{\theta}}{p_0 + i v_F \abs{\bm{p}} \cos{\theta}}\,.
\end{equation}
Integrating next the product of $G_{k+q/2} G_{k-q/2}$ over $\varepsilon_{\bm{k}}$ and $k_0$, where $k_0$ is the fermionic Matsubara frequency, and multiplying the result by $\Gamma (p)$, we obtain the Landau-damping term in a dressed polarization bubble as
\begin{equation}
    \Pi_0(p)  = \int \frac{d\theta}{2\pi} \Gamma (p) \frac{p_0}{p_0 +i Z v_F \abs{\bm{p}}\cos \theta} =
    \Pi^{(0)}(p) \,.
\end{equation}
Therefore, the Landau-damping term in the bosonic propagator remains the same as for free fermions \textit{for arbitrary} ratio of $\abs{p_0}/v_F \abs{\bm{p}}$.

We now combine these two pieces of information --- the absence of renormalization of $\Pi_0(p)$ and the relation $\Gamma_l = 1+\lambda =1/Z$ --- to reevaluate the polarization bubble using dressed Green's functions and dressed vertices. For even $l$, a straightforward extension of the analysis in the previous section yields
\begin{equation}
  \Pi_{l} (\omega) \approx \Pi^{Z=1}_{l} (\omega)  Z^2 \Gamma_l^2
\end{equation}
Using $\Gamma_l \approx 1/Z$, we find that $\Pi_l (\omega)$  remains the same as for free fermions and bare vertices.

For odd $l$, the analysis is a bit more tricky (see Appendix~\ref{Sec:appendix}), but  the final result is the same --- $\Pi_l (\omega)$ remains the same as for free fermions and bare vertices.

We did not extend the analysis of the residue and vertex corrections to the NFL regime, but given that we found the cancellation for any value of $Z$, no matter how small, we expect that the cancellation also holds in the quantum-critical regime, where $Z =0$.

The issue of stability of collective excitations near a QCP has been recently analyzed by H.\ Guo~\cite{Guo}. He argued that for the specific vertex $\mathcal{V}_{\!l} (\bm{k}) = \cos(l \theta_{\bm{k} \hat{x}}) \Psi_l (\abs{\bm{k}})$ with $\Psi_l (\abs{\bm{k}}) = 1 + l^2(\abs{\bm{k}}/k_F -1)$, which reduces to a momentum vertex for $l=1$, the contribution to $\Pi (\omega)$ that we obtained cancels out. In the Fermi liquid regime where $Z \approx 1$, he obtained $\Pi_l (\omega)$ with the same functional form as in~\eqref{nn_1} but with different sign of the $\omega^2$ term and without a logarithm for the $\omega^3$ term. Also, his prefactor contains $l^2 (l^2-1)^2$ instead of $l^4$ and additional $\xi^{-2}$. He further argued that vertex renormalizations enhance his $\Pi_l (\omega)$ by $1/Z^2 \propto \xi^2$, such that it becomes comparable to the one in~\eqref{nn_1}. However, because of a different sign of the $\omega^2$ term, the extension of his $\Pi_l (\omega)$ to the non-Fermi liquid regime yields unstable excitations that break causality. We reproduced Guo's results for a Fermi liquid. We didn't explicitly compute self-energy and vertex corrections for his form of $\mathcal{V}_{\!l} (\bm{k})$, but given that we found an explicit cancellation between self-energy and vertex corrections in our case for both even and odd $l$, we conjecture that they will also cancel out for his form of $\mathcal{V}_{\!l} (\bm{k})$. Without such corrections, the full susceptibility in Eq.~\eqref{1_c} remains causal for all $l$ and at any distance from a QCP\@.

\textit{Conclusions.}~
In this communication, we presented a study of the spectral properties of collective excitations with even and odd angular momentum $l >0$ in both the FL and NFL regimes near a QCP, using a charge Pomeranchuk instability with $l=0$ as an example. Previous works on interaction-driven damping rates of zero-sound modes in a FL found a sharp distinction  between the modes with even and odd $l$. Our objective was to extend this analysis into the NFL regime, including the QCP itself, and to address a recently raised question whether causality survives in a NFL\@. To this end, we derived the full crossover function describing the damping rate across FL and NFL regimes. Our results demonstrate that causality is not broken: the NFL state near a QCP, and even at the QCP itself, is stable with respect to fluctuations in all angular momentum channels different from the critical one.

\textit{Acknowledgments.}~
We acknowledge with thanks useful discussions with R.~Fernandes, H.~Goldman, A.~Kamenev, A.~Klein, A.~Levchenko, L.~Levitov, P.~Nosov, S.~Sachdev, J.~Schmalian and especially H.~Guo.  The work of Y.G.\ was supported by the Simons Foundation Grant No.~1249376;  A.V.Ch.\ was supported the by U.S.\ Department of Energy, Office of Science, Basic Energy Sciences, under Award No.~DE-SC0014402;  D.L.M.\ was supported by the NSF grant DMR-2224000. A.V.C.\ and D.L.M.\ acknowledge the hospitality of the Aspen Center for Physics, where some of the work on this project has been performed. The Aspen Center for Physics is supported by NSF Grant No.~PHY-2210452. D.L.M.\ acknowledges support for his stay at the University of Minnesota to work on this project from the Simons Foundation Targeted Grant 920184 to the Fine Theoretical Physics Institute.

\appendix
\section{Derivation of the expression for $\Pi_l (\omega)$}
\label{Sec:appendix}
In this section we perform a diagrammatic analysis of the polarization bubble with angular momentum $l$, paying special attention to the case of odd channels.  We  do calculations on the Matsubara axis, obtain analytical expressions for $\Pi_l (q_0)$, where $q_0$ is a bosonic Matsubara frequency, and then rotate to the real axis.

A propagator of a free electron on the Matsubara axis is
\begin{equation}
  \label{eq:el_propagator}
  G_{k} = \frac{1}{i k_0 - \varepsilon_k}\,,
\end{equation}
where $\varepsilon_{k}$ includes the chemical potential. The contributions from Fig.~1 of the main text are studied separately in the forthcoming sections.

\subsection{Self-energy and Maki-Thompson contributions}
\label{Sec:MT}

\begin{figure}[htb]
  \centering
  \subfloat[]{%
   \includegraphics[width=0.5\linewidth]{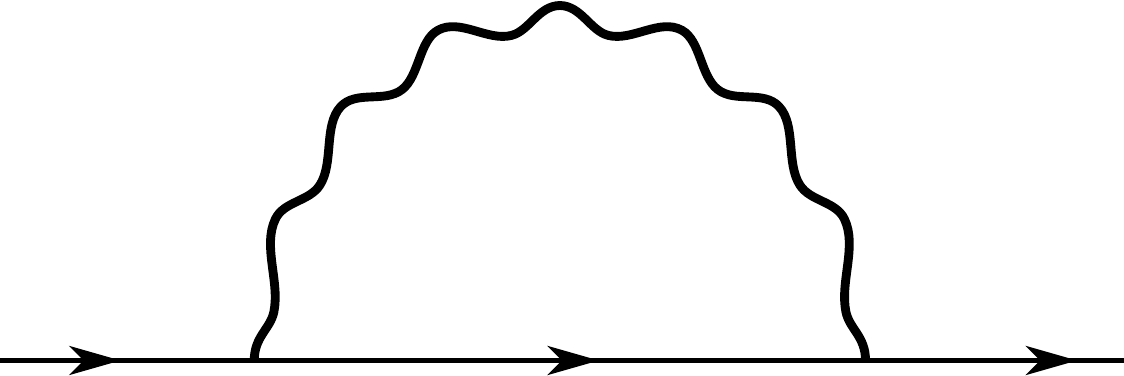}  
  \label{fig:sfe}
  }\hfill
  \subfloat[]{%
    \includegraphics[width=0.4\linewidth]{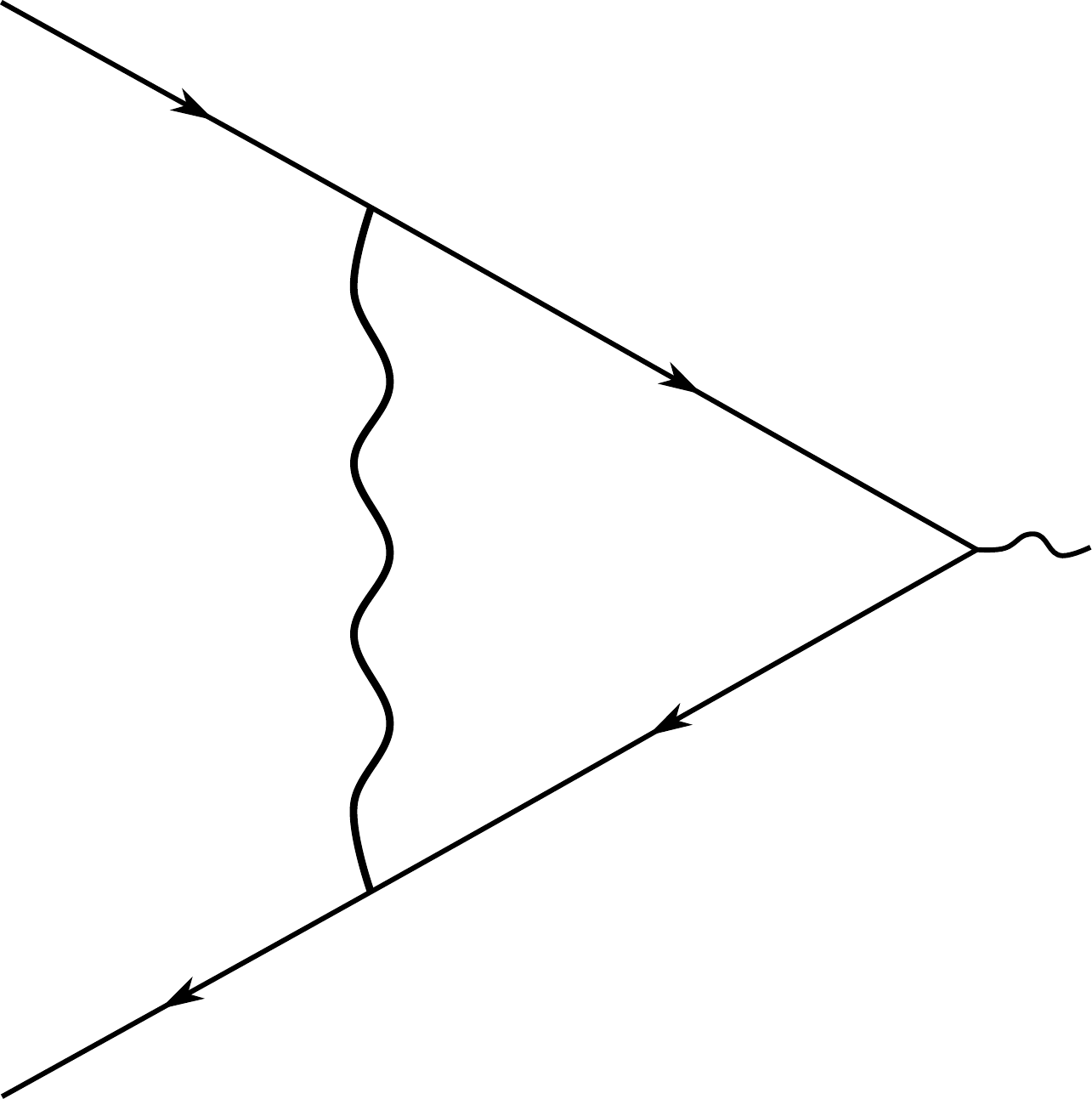}%
    \label{fig:sv}%
  }
  \caption{The lowest order diagrams for the self-energy (a) and  vertex renormalization (b).}
\end{figure}

In this section we focus on the SE+MT contributions to $\Pi_l (\omega)$ polarization bubble.

The SE diagrams in Figs.~1a and~1b of the main text give, respectively,
\begin{equation}
    \Pi_{l, SE_1} (q_0) = - \int \frac{d^3 k}{{(2 \pi)}^3} G^2_{k+q}G_{k} \Sigma_{k+q} \,{(\mathcal{V}_{\!l} (\bm{k}))}^2
\end{equation}
and
\begin{equation}
    \Pi_{l, SE_2} (q_0) = - \int \frac{d^3 k}{{(2 \pi)}^3} G_{k+q}G^2_{k} \Sigma_{k} \, {(\mathcal{V}_{\!l} (\bm{k}))}^2\,,
\end{equation}
where the self-energy part in Fig.~\ref{fig:sfe} is
\begin{equation}
    \Sigma_{k} = \frac{\bar{g}}{\chi_0} \int \frac{d^3 p}{{(2\pi)}^3} \chi (p) G_{k+p}\,.
\end{equation}

Using the free-electron propagator identity
\begin{equation}
  G_{k}G_{k+q} = \frac{G_{k} - G_{k+q}}{iq_0}\,,
\end{equation}
one obtains
\begin{equation}
  \begin{split}
  \label{eq:cancel_1}
      & \Pi_{l, SE_1} (q_0) + \Pi_{l, SE_2}(q_0) = - \frac{1}{i q_0} \int \frac{d^3 k}{{(2 \pi)}^3} \Big\{ \left(G^2_{k}\Sigma_{k}
      - G^2_{k+q} \Sigma_{k+q}\right)\\
      & + {}G_{k}G_{k+q} \left(\Sigma_{k+q} - \Sigma_{k}\right)\Big\} {(\mathcal{V}_{\!l} (\bm{k}))}^2\\
      =& - \frac{1}{i q_0} \int \frac{d^3 k}{{(2 \pi)}^3}  G_{k}G_{k+q} \left(\Sigma_{k+q} - \Sigma_{k}\right)
      {(\mathcal{V}_{\!l} (\bm{k}))}^2\,,
  \end{split}
\end{equation}
where the first term in parentheses vanishes upon shifting the integration variable $k+q \to k$.

The MT diagram in Fig.~1c of the main text is
\begin{align}
   & \Pi_{l, MT} (q_0)  = \\
   & -\frac{\bar{g}}{\chi_0} \int \frac{d^3 k d^3 p}{{(2\pi)}^6} \chi(p)
     G_{k}G_{k+q}G_{k+p}G_{k+p+q}
    \, \mathcal{V}_{\!l} (\bm{k}) \mathcal{V}_{\!l} (\bm{k} + \bm{p})\,,\nonumber
\end{align}
which we split into two parts as $ \Pi_{l, MT} (q_0) \equiv \Pi_{_l,MT1} (q_0) + \Pi_{l,MT_2} (q_0)$. The first part is
\begin{equation}
    \begin{split}
        \Pi_{l, MT_1} (q_0) &= -\frac{\bar{g}}{\chi_0} \int \frac{d^3 k d^3 p}{{(2\pi)}^6} \chi(p) G_{k}G_{k+q}G_{k+p}G_{k+p+q}\,  {(\mathcal{V}_{\!l} (\bm{k}))}^2\\
        &= - \int \frac{d^3 k}{{(2\pi)}^3} G_{k}G_{k+q} \Gamma(k+q;q)\,
         {(\mathcal{V}_{\!l} (\bm{k}))}^2 \,,
    \end{split}
\end{equation}
where the scalar vertex $\Gamma(k+q;q)$, shown in Fig.~\ref{fig:sv}, is
\begin{equation}
    \Gamma(k+q;q) = \frac{\bar{g}}{\chi_0} \int \frac{d^3 p}{{(2\pi)}^3} \chi(p)
     G_{k+p} G_{k+p+q}\,.
\end{equation}
It satisfies the Ward identity
\begin{equation}
     \Gamma(k+q;q) = \frac{\Sigma_{k} - \Sigma_{k+q}}{i q_0}\,.
\end{equation}
Hence
\begin{equation}
    \begin{split}
        \Pi_{l, MT_1} (q_0) &= - \int \frac{d^3 k}{{(2\pi)}^3}  G_{k}G_{k+q} \frac{\Sigma_{k} - \Sigma_{k+q}}{i q_0}\,  {(\mathcal{V}_{\!l} (\bm{k}))}^2 \\
        &= - \left(\Pi_{SE_1} (q_0) + \Pi_{SE_2} (q_0)\right)\,.
    \end{split}
\end{equation}
Thus, $\Pi_{l, MT_1} (q_0)+\Pi_{l, SE_1} (q_0) + \Pi_{l, SE_2} (q_0)=0$, and the net contribution from the MT+SE diagrams is
\begin{widetext}
\begin{align}
    \label{Eq:app-ref-1}
      \Pi_{l, MT+SE} (q_0)& = \Pi_{l, MT_2} (q_0)
      = -\frac{\bar{g}}{\chi_0} \int \frac{d^3 k d^3 p}{{(2\pi)}^6} \chi(p) G_{k}G_{k+q}G_{k+p}G_{k+p+q}\,
      \mathcal{V}_{\!l} (\bm{k})
      \left(\mathcal{V}_{\!l}(\bm{k} + \bm{p}) -\mathcal{V}_{\!l} (\bm{k})\right)\\
      &= \frac{\bar{g}}{\chi_0 q_0^2}\int \frac{d^3 k d^3 p}{{(2\pi)}^6} \chi(p) \mathcal{V}_{\!l} (\bm{k}-\tfrac{\bm{p}}{2})  \left(\mathcal{V}_{\!l} (\bm{k} + \tfrac{\bm{p}}{2}) -\mathcal{V}_{\!l} (\bm{k}-\tfrac{\bm{p}}{2})\right)
      \left\{2G_{k-\tfrac{p}{2}}G_{k+\tfrac{p}{2}} - G_{k-\tfrac{p}{2}}G_{k+\tfrac{p}{2}+q} - G_{k+\tfrac{p}{2}}G_{k-\tfrac{p}{2}+q}\right\} \,.\notag
\end{align}
\end{widetext}
Because the free-fermion propagator has a simple pole structure, the fermion frequency integral is straightforward:
\begin{widetext}
  \begin{equation}
    \begin{split}
      &\int \frac{dk_0}{2\pi} \Big\{2G_{k-\tfrac{p}{2}}G_{k+\tfrac{p}{2}} - G_{k-\tfrac{p}{2}}G_{k+\tfrac{p}{2}+q} - G_{k+\tfrac{p}{2}}G_{k-\tfrac{p}{2}+q}\Big\}=\\
      &=\left(f_{\bm{k}+\tfrac{\bm{p}}{2}}-f_{\bm{k}-\tfrac{\bm{p}}{2}}\right)\left[\frac{1}{i(p_0+q_0)+\varepsilon_{\bm{k}-\tfrac{\bm{p}}{2}}-\varepsilon_{\bm{k}+\tfrac{\bm{p}}{2}}}+\frac{1}{i(p_0-q_0)+\varepsilon_{\bm{k}-\tfrac{\bm{p}}{2}}-\varepsilon_{\bm{k}+\tfrac{\bm{p}}{2}}}-\frac{2}{i p_0+\varepsilon_{\bm{k}-\tfrac{\bm{p}}{2}}-\varepsilon_{\bm{k}+\tfrac{\bm{p}}{2}}}\right]\,,
    \end{split}
  \end{equation}
\end{widetext}
where $f_{\bm{k}}\equiv\Theta (-\varepsilon_{\bm{k}})$ is the equilibrium distribution function at $T=0$. Expanding to leading order in the boson momentum $\bm{p}$ and switching to polar coordinates via
\begin{equation}
  \grad_{\bm{k}} = \frac{\hat{e}_{\theta}}{k}\frac{\partial}{\partial\theta}+\hat{e}_{k}\frac{\partial}{\partial k}\,,  
\end{equation}
we find
\begin{equation}
\label{Eq:Fermi_expansion}
  \begin{split}
    f_{\bm{k}+\tfrac{\bm{p}}{2}}-f_{\bm{k}-\tfrac{\bm{p}}{2}} &= \bm{p} \cdot \grad_{\bm{k}}f=-(\bm{p}\cdot \hat{e}_{k})\delta(\abs{\bm{k}}-k_F)\\
    &=-\abs{\bm{p}} \cos \theta_{\bm{k}\bm{p}}\delta(\abs{\bm{k}}-k_F)\,,
  \end{split}
\end{equation}
where $\theta_{\bm{k}\bm{p}}\equiv \theta_{\bm{k}\hat{x}}-\theta_{\bm{p}\hat{x}}$ is the angle between $\bm{k}$ and $\bm{p}$.

Similarly,
\begin{align}
  \label{Eq:mt_vertex_expansion}
    \mathcal{V}_{\!l} (\bm{k}-\tfrac{\bm{p}}{2})  \big(\mathcal{V}_{\!l} (\bm{k} + \tfrac{\bm{p}}{2}) &-\mathcal{V}_{\!l} (\bm{k}-\tfrac{\bm{p}}{2})\big)\\
        &= \frac{l \abs{\bm{p}}}{2\abs{\bm{k}}}\sin \theta_{\bm{k}\bm{p}}\sin 2l(\theta_{\bm{k}\bm{p}}+\theta_{\bm{p}\hat{x}})\notag\\
        &-\frac{l^2\bm{p}^2}{2\bm{k}^2}\sin^2\theta_{\bm{k}\bm{p}}\sin^2 l(\theta_{\bm{k}\bm{p}}+\theta_{\bm{p}\hat{x}})\,.\notag
\end{align}
Upon integrating the last equation over $\theta_{\bm{p}\hat{x}}$, the first term vanishes, and $\sin^2 l(\theta_{\bm{k}\bm{p}}+\theta_{\bm{p}\hat{x}})$ is replaced by its average value of $1/2$.

As a result, Eq.~\eqref{Eq:app-ref-1} simplifies to
\begin{equation}
\label{Eq:pimtse}
  \Pi_{l, MT+SE} (q_0) =\frac{\bar{g} l^2}{q_0^2 2^5 \pi^3 k_F v_F}\int \frac{\abs{\bm{p}}^3 d\abs{\bm{p}} dp_0}{\bm{p}^2 + \xi^{-2}+ \gamma \tfrac{v_F\abs{p_0}}{\sqrt{v_F^2\bm{p}^2+p_0^2}}} \mathcal{P}_l(p,q_0)\,,
\end{equation}
where
\begin{align}
\label{Eq:pol_ferm}
  &\mathcal{P}_l(p,q_0) = \\
  &\int \frac{d\theta}{2\pi} \cos \theta \sin^2\theta\left[\frac{1}{is_+-\cos \theta}+\frac{1}{is_--\cos \theta}-\frac{2}{is_0-\cos \theta}\right]\,\notag
\end{align}
and
\begin{equation}
  s_{\pm}=\frac{p_0\pm q_0}{\abs{\bm{p}} v_F},\;s_{0}=\frac{p_0}{\abs{\bm{p}} v_F}\,.
\end{equation}
Carrying out the angular integration yields
\begin{align}
  \mathcal{P}_l (p,q_0) ={}& \abs{s_+}\left(\sqrt{s_+^2+1}-\abs{s_+}\right)+\abs{s_-}\left(\sqrt{s_-^2+1}-\abs{s_-}\right)\notag\\
  &-2\abs{s_0}\left(\sqrt{s_0^2+1}-\abs{s_0}\right)\,.
\end{align}
The combination $2s_0^2-s_+^2-s_-^2$ gives rise to an unimportant constant shift in $\Pi_{l,MT+SE} (q_0)$. Expanding the remaining terms in small $(s_{\pm},s_{0})$ gives
\begin{equation}
\label{Eq:ferm_expansion}
  \mathcal{P}_l (p,q_0) =(\abs{s_+} +\abs{s_-} -2\abs{s_0})+\frac12 (\abs{s_+}^3+\abs{s_-}^3-2\abs{s_0}^3)\,.
\end{equation}
The two terms represent, respectively, the leading ($\sim \abs{s}$) and subleading ($\sim \abs{s}^3$) contributions to $\mathcal{P}_l (p,q_0)$.

Further calculation of the polarization bubble in Eq.~\eqref{Eq:pimtse} depends on how close the system is to the QCP\@.

\subsubsection{Fermi liquid regime}
In the FL regime, Landau damping is weak, and the bosonic propagator can be expanded in $p_0/v_F \abs{\bm{p}}$:
\begin{equation}
\label{Eq:bos_expansion}
  \begin{split}
    \chi (p)=&-\frac{\chi_0}{{(\bm{p}^2+\xi^{-2})}^2}\frac{\gamma v_F p_0}{\sqrt{v_F^2\bm{p}^2+p_0^2}}\\
    =& -\frac{\chi_0}{{(\bm{p}^2+\xi^{-2})}^2}\frac{\gamma p_0}{\abs{\bm{p}}}\left(1-\frac{p_0^2}{2 v_F^2 \bm{p}^2}+\ldots\right)\,,
  \end{split}
\end{equation}
where the two terms of the expansion in braces represent, respectively, the leading and subleading contributions to $\chi (p)$.

The leading contribution to the polarization bubble on the Matsubara axis is given by
\begin{equation}
\label{Eq:MT+SE}
  \begin{split}
    &\Pi_{l, MT+SE} (q_0) =\\
     &={} -\frac{\bar{g} l^2}{q_0^2 2^5 \pi^3 k_F v_F}\int \frac{\abs{\bm{p}}^3 d\abs{\bm{p}} dp_0}{{(\bm{p}^2+\xi^{-2})}^2}\frac{\gamma p_0}{\abs{\bm{p}}}(\abs{s_+} +\abs{s_-} -2\abs{s_0})\\
    &={} -q_0 \bar{g}^2 \frac{l^2 \xi^2}{192 \pi^4 v_F^4}\,.
  \end{split}
\end{equation}
Rotating to real frequencies $i q_0 \to \omega +i0$, we obtain
\begin{equation}
\label{Eq:MT+SE_re}
  \Pi_{l, MT+SE} (\omega)= i\omega \bar{g}^2 \frac{l^2 \xi^2}{192 \pi^4 v_F^4}\,.
\end{equation}

Subleading corrections from the fermion polarization function in Eq.~\eqref{Eq:ferm_expansion} and the bosonic propagator in Eq.~\eqref{Eq:bos_expansion} generate terms of order $\propto \bar{g}^2 l^2 q_0^3 \xi^4 \ln \xi^3 q_0$. One can show that these contributions cancel each other exactly, leaving Eq.~\eqref{Eq:MT+SE} as the final result for the MT+SE diagrams in the FL regime, to order $\mathcal{O}(\omega^3)$.

\subsubsection{Quantum Critical Point}
Here, we focus on the SE + MT contribution to the polarization bubble exactly at QCP, when the correlation length can be put to infinity, $\xi \to \infty$. Combining Eq.~\eqref{Eq:pimtse} and Eq.~\eqref{Eq:ferm_expansion}, one gets for the leading contribution:
\begin{equation}
\label{Eq:pimtse_qcp}
  \Pi_{l, MT+SE}=\frac{\bar{g} l^2}{q_0^2 2^3 \pi^3 k_F v_F^2}\int_0^{\infty}\!\! d\abs{\bm{p}}\, \abs{\bm{p}}^3 \int_0^{q_0} \!\! dp_0 \frac{q_0-p_0}{\abs{\bm{p}}^3 + \gamma p_0}\,.
\end{equation}
The frequency integral equals
\begin{align}
  \int_0^{q_0} dp_0 \frac{q_0-p_0}{\abs{\bm{p}}^3 + \gamma p_0} = \frac{1}{\gamma^2}(\abs{\bm{p}}^3 + \gamma q_0)\ln \frac{\abs{\bm{p}}^3 + \gamma q_0}{\abs{\bm{p}}^3}-\frac{q_0}{\gamma}\,.
\end{align}
Its asymptotic form $\sim \tfrac{q_0^2}{2\abs{\bm{p}}^3}$ at $\abs{\bm{p}} \to \infty$ gives rise to a static, pure real, contribution to the bubble in Eq.~\eqref{Eq:pimtse_qcp}, which is of no interest to us. Subtracting this contribution as a counter-term, we get for the dynamic part of the bubble
\begin{widetext}
  \begin{equation}
  \label{Eq:pimtse_qcp1}
    \begin{split}
      \Pi_{l, MT+SE}&={}\frac{\bar{g} l^2}{q_0^2 2^3 \pi^3 k_F v_F^2}\int_0^{\infty}d\abs{\bm{p}}\, \abs{\bm{p}}^3 \left(\frac{1}{\gamma^2}(\abs{\bm{p}}^3 + \gamma q_0)\ln \frac{\abs{\bm{p}}^3 + \gamma q_0}{\abs{\bm{p}}^3}-\frac{q_0}{\gamma}-\frac{q_0^2}{2\abs{\bm{p}}^3}{}\right)\\
      &={}
      \frac{\bar{g} l^2 q_0^{\frac13}\gamma^{\frac13}}{{(2\pi)}^3 k_F v_F^2}\int_0^{\infty}d\abs{\bm{p}}\, \abs{\bm{p}}^3 \left((\abs{\bm{p}}^3 + 1)\ln \frac{\abs{\bm{p}}^3 + 1}{\abs{\bm{p}}^3}-1-\frac{1}{2\abs{\bm{p}}^3}\right)\,,
    \end{split}
  \end{equation}
\end{widetext}
where in the transition from the first to the second line, the momentum has been re-scaled as $\abs{\bm{p}}\to \abs{\bm{p}} q_0^{\frac13}\gamma^{\frac13}$. The integral in the second line equals $-\sqrt{3}\pi/14$, so we end up with
\begin{equation}
\label{Eq:pimtse_qcp2}
    \Pi_{l, MT+SE}=
    -\frac{\bar{g}^{\frac43} l^2 q_0^{\frac13}\sqrt{3}}{7{\pi}^{\frac{7}{3}}2^{\frac{13}{3}} v_F^{\frac{10}{3}}m^{\frac23}}\,.
\end{equation}
This result is valid both for even and odd $l$.

\subsection{Aslamazov-Larkin contribution}
\label{Sec:AL}

In explicit form, the two  AL diagrams in Figs.~1d and~1e of the main text yield
 \begin{widetext}
  \begin{equation}
    \begin{split}
      \Pi_{l, AL_1} &={} \frac{\bar{g}^2}{\chi_0^2} \int \frac{d^3 k d^3 k' d^3 p}{{(2\pi)}^9}
      G_{k-\tfrac{p}{2}-\tfrac{q}{2}}G_{k-\tfrac{p}{2}+\tfrac{q}{2}}G_{k+\tfrac{p}{2}}G_{k'+\tfrac{p}{2}+\tfrac{q}{2}}G_{k'+\tfrac{p}{2}-\tfrac{q}{2}}G_{k'-\tfrac{p}{2}} \chi(p-\tfrac{q}{2})\chi(p+\tfrac{q}{2}) \mathcal{V}_{\!l} (\bm{k}-\tfrac{\bm{p}}{2})  \mathcal{V}_{\!l}(\bm{k}'+\tfrac{\bm{p}}{2})\,,\\
      \Pi_{l, AL_2} &={} \frac{\bar{g}^2}{\chi_0^2} \int \frac{d^3 k d^3 k' d^3 p}{{(2\pi)}^9}
      G_{k-\tfrac{p}{2}-\tfrac{q}{2}}G_{k-\tfrac{p}{2}+\tfrac{q}{2}}G_{k+\tfrac{p}{2}}G_{k'-\tfrac{p}{2}+\tfrac{q}{2}}G_{k'-\tfrac{p}{2}-\tfrac{q}{2}}G_{k'+\tfrac{p}{2}} \chi(p-\tfrac{q}{2})\chi(p+\tfrac{q}{2}) \mathcal{V}_{\!l} (\bm{k}-\tfrac{\bm{p}}{2})  \mathcal{V}_{\!l} (\bm{k}'-\tfrac{\bm{p}}{2})\,.
    \end{split}
  \end{equation}
For these diagrams, the integrals over fermionic frequencies differ only in sign,
  \begin{equation}
  \label{Eq:freq_int}
    \begin{split}
      &\int \frac{dk_0 dk'_0}{{(2\pi)}^2}
      G_{k-\tfrac{p}{2}-\tfrac{q}{2}}G_{k-\tfrac{p}{2}+\tfrac{q}{2}}G_{k+\tfrac{p}{2}}G_{k'\pm\tfrac{p}{2}+
      \tfrac{q}{2}}G_{k'\pm\tfrac{p}{2}-\tfrac{q}{2}}G_{k'\mp\tfrac{p}{2}}=\\
      ={}&\pm \frac{1}{{(iq_0)}^2}\left(f_{\bm{k}+\tfrac{\bm{p}}{2}}-f_{\bm{k}-\tfrac{\bm{p}}{2}}\right)
      \left[\frac{1}{i(p_0+\tfrac{q_0}{2})+\varepsilon_{\bm{k}-\tfrac{\bm{p}}{2}}-\varepsilon_{\bm{k}+\tfrac{\bm{p}}{2}}}-
      \frac{1}{i(p_0-\tfrac{q_0}{2})+\varepsilon_{\bm{k}-\tfrac{\bm{p}}{2}}-\varepsilon_{\bm{k}+\tfrac{\bm{p}}{2}}}\right]\\
      &{}\times \left(f_{\bm{k}'+\tfrac{\bm{p}}{2}}-f_{\bm{k}'-\tfrac{\bm{p}}{2}}\right)
      \left[\frac{1}{i(p_0+\tfrac{q_0}{2})+\varepsilon_{\bm{k}'-\tfrac{\bm{p}}{2}}-
      \varepsilon_{\bm{k}'+\tfrac{\bm{p}}{2}}}-\frac{1}{i(p_0-\tfrac{q_0}{2})+
      \varepsilon_{\bm{k}'-\tfrac{\bm{p}}{2}}-\varepsilon_{\bm{k}'+\tfrac{\bm{p}}{2}}}\right]\,.
    \end{split}
  \end{equation}
 In the sum $\Pi_{l, AL_1}+\Pi_{l, AL_2}\equiv \Pi_{l, AL}$,
 the vertices are combined into
  \begin{equation}
      \mathcal{V}_{\!l} (\bm{k}-\tfrac{\bm{p}}{2})  \big(\mathcal{V}_{\!l} (\bm{k}' + \tfrac{\bm{p}}{2}) -\mathcal{V}_{\!l} (\bm{k}'-\tfrac{\bm{p}}{2})\big) = \frac{l\abs{\bm{p}}}{\abs{\bm{k'}}}\cos l\theta_{\bm{k}\hat{x}} \sin\theta_{\bm{k}'\bm{p}}\sin l\theta_{\bm{k}'\hat{x}}
      -\frac{l^2\bm{p}^2}{2\abs{\bm{k}} \abs{\bm{k'}}}\sin\theta_{\bm{k}\bm{p}} \sin l\theta_{\bm{k}\hat{x}}\sin\theta_{\bm{k}'\bm{p}} \sin l\theta_{\bm{k}'\hat{x}}\,.
  \end{equation}
  The first term leads to a contribution that is odd in $\bm{p}$, which integrates to zero over $\bm{p}$ direction, by symmetry. The second term can be transformed as follows,
  \begin{align}
  \label{Eq:vertex_expansion}
    &-\frac{l^2\bm{p}^2}{2\abs{\bm{k}} \abs{\bm{k'}}}\sin\theta_{\bm{k}\bm{p}} \sin l\theta_{\bm{k}\hat{x}}\sin\theta_{\bm{k}'\bm{p}} \sin l\theta_{\bm{k}'\hat{x}}
    =-\frac{l^2\bm{p}^2}{4\abs{\bm{k}} \abs{\bm{k'}}}\sin\theta_{\bm{k}\bm{p}}\sin\theta_{\bm{k}'\bm{p}} \left(\cos l(\theta_{\bm{k}\bm{p}}-\theta_{\bm{k}'\bm{p}})-\cancel{\cos l(\theta_{\bm{k}\bm{p}}+\theta_{\bm{k}'\bm{p}}+2\theta_{\bm{p}\hat{x}})}\right)\to \\
    &{}-\frac{l^2\bm{p}^2}{4\abs{\bm{k}} \abs{\bm{k'}}}\left(\cancel{\sin\theta_{\bm{k}\bm{p}}\cos l\theta_{\bm{k}\bm{p}}\sin\theta_{\bm{k}'\bm{p}}\cos l\theta_{\bm{k}'\bm{p}}}+\sin\theta_{\bm{k}\bm{p}}\sin l\theta_{\bm{k}\bm{p}}\sin\theta_{\bm{k}'\bm{p}}\sin l\theta_{\bm{k}'\bm{p}}\right)\to -\frac{l^2\bm{p}^2}{4\abs{\bm{k}} \abs{\bm{k'}}}\sin\theta_{\bm{k}\bm{p}}\sin l\theta_{\bm{k}\bm{p}}\sin\theta_{\bm{k}'\bm{p}}\sin l\theta_{\bm{k}'\bm{p}}\,,\notag
  \end{align}
\end{widetext}
since $\cos l(\theta_{\bm{k}\bm{p}}+\theta_{\bm{k}'\bm{p}}+2\theta_{\bm{p}\hat{x}})$ gives zero when integrated over $\theta_{\bm{p}\hat{x}}$, while the contribution from the first term in the second line, $\sin\theta_{\bm{k}\bm{p}}\cos l\theta_{\bm{k}\bm{p}}\sin\theta_{\bm{k}'\bm{p}}\cos l\theta_{\bm{k}'\bm{p}}$, integrates to zero over $\theta_{\bm{k}\bm{p}}$ and $\theta_{\bm{k}'\bm{p}}$.

Using Eqs.~\eqref{Eq:Fermi_expansion},~\eqref{Eq:freq_int} and~\eqref{Eq:vertex_expansion}, we obtain for the AL contribution to the polarization bubble
\begin{equation}
\label{Eq:piAL}
  \Pi_{l, AL} = -\frac{\bar{g}^2l^2}{64\pi^2\chi_0^2q_0^2 v_F^2}\int \abs{\bm{p}}^3 d\abs{\bm{p}}\,dp_0 \chi(p-\tfrac{q}{2})\chi(p+\tfrac{q}{2})\mathcal{P}^2_l (p,q_0),
\end{equation}
where
\begin{align}
\label{Eq:pol_ferm_al}
  &\mathcal{P}_l(p,q_0) = \notag\\
  & \int \frac{d\theta}{2\pi} \cos \theta \sin\theta \sin l\theta\left[\frac{1}{i s_+ -\cos \theta} - \frac{1}{i s_- -\cos \theta}\right]=\notag\\
  &={} i e^{-i\frac{\pi}{2} l}\Big[\abs{s_+}{(\sgn{s_+})}^{l-1}{\left(\sqrt{s_+^2+1}-\abs{s_+}\right)}^l\\
  &-{}\abs{s_-}{(\sgn{s_-})}^{l-1}{\left(\sqrt{s_-^2+1}-\abs{s_-}\right)}^l\Big]\, \notag
\end{align}
with
\begin{equation}
  s_{\pm}=\frac{p_0\pm \tfrac{q_0}{2}}{\abs{\bm{p}} v_F}\,.
\end{equation}

Consider first even $l$ channels. Expanding the bubble in Eq.~\eqref{Eq:pol_ferm_al} in powers of small $s_{\pm}$, we obtain for the leading term of this expansion
\begin{equation}
  \mathcal{P}_l (p,q_0)\sim i e^{i\pi n}(s_{+}-s_{-})=i e^{i\pi n}\frac{q_0}{v_F \abs{\bm{p}}}\,.
\end{equation}
Substituting the last result into Eq.~\eqref{Eq:piAL}, we obtain a frequency-independent term, $\Pi_{AL} \propto \bar{g}^2 l^2 \xi^2$, which does not contribute to the imaginary part of the bubble, $\Pi''$. This leaves the MT term of Eq.~\eqref{Eq:MT+SE_re} as the leading contribution to the imaginary part of the polarization bubble for even $l$.

For odd $l$ the situation changes qualitatively: the leading contributions from AL and MT diagrams cancel each other exactly to leave the AL subleading term as the final result. We demonstrate this below in both the
FL and NFL regimes.

The expansion of the bubble in powers of small $s_{\pm}$ is now given by
\begin{equation}
\label{Eq:al_expansion}
  \begin{split}
    \mathcal{P}^2_l (p,q_0) &\sim {}{(\abs{s_{+}}-\abs{s_{-}})}^2 -2l {(\abs{s_{+}}-\abs{s_{-}})}^2(\abs{s_{+}}+\abs{s_{-}})\\
    &+{}l^2 {(\abs{s_{+}}-\abs{s_{-}})}^2 (2s_{+}^2+3\abs{s_{+}}\abs{s_{-}}+2s_{-}^2)\,.
  \end{split}
\end{equation}
One can show that the second, linear in $l$, term of this expansion does not contribute to $\Pi''$, and for this reason is irrelevant for our purposes. In the following subsections, we will focus on the first and third terms of the expansion, which give, respectively, the leading and subleading contributions to the dissipative part of the polarization.

\subsubsection{Cancellation of leading contributions in odd channels}
\begin{widetext}
In the FL regime, the leading contribution to the polarization bubble from the AL diagrams is given by~\footnote{The frequency integral is regularized by the regular (i.e., $\xi$-independent)  counter-term $q_0^2/(v_F^2 \bm{p}^2)$.}
\begin{equation}
    \Pi_{l, AL} = -\frac{\bar{g}^2l^2}{64 \pi^4 q_0^2 v_F^2}\int \frac{\abs{\bm{p}}^3 d\abs{\bm{p}}\,dp_0}{{(\bm{p}^2+\xi^{-2})}^2} [{(\abs{s_{+}}-\abs{s_{-}})}^2-\tfrac{q_0^2}{v_F^2 \bm{p}^2}]
    = q_0 \bar{g}^2 \frac{l^2 \xi^2}{192 \pi^4 v_F^4}\,.
\end{equation}
This expression cancels out the MT+SE contribution from Eq.~\eqref{Eq:MT+SE}. This cancellation holds at the QCP as well. In this case,
  \begin{equation}
    \begin{split}
      \Pi_{l, AL} &={} -\frac{\bar{g}^2l^2}{64 \pi^4 q_0^2 v_F^2}\int \frac{\abs{\bm{p}}^3 d\abs{\bm{p}}\,dp_0}{\left(\bm{p}^2+\gamma \frac{\abs{p_0-\tfrac{q_0}{2}}}{\abs{\bm{p}}}\right)\left(\bm{p}^2+\gamma \frac{\abs{p_0+\tfrac{q_0}{2}}}{\abs{\bm{p}}}\right)} {(\abs{s_{+}}-\abs{s_{-}})}^2\\
      &={} -\frac{\bar{g}^2l^2}{16 \pi^4 q_0^2 v_F^4\gamma^3}\int d\abs{\bm{p}}\, \abs{\bm{p}}^3\left[(\abs{\bm{p}}^3+\gamma q_0) \ln \left(1+\frac{\gamma q_0}{\abs{\bm{p}}^3}\right)-\gamma q_0 -\frac{\gamma^2 q_0^2}{2\abs{\bm{p}}^3}\right]\,,
    \end{split}
  \end{equation}
where the momentum integral has been regularized by adding a regular counter-term $-\gamma^2 q_0^2/(2 \abs{\bm{p}}^3)$. Re-scaling the momentum via $\abs{\bm{p}} \to \abs{\bm{p}} q_0^{\frac13} \gamma^{\frac13}$, we get
  \begin{equation}
    \begin{split}
      \Pi_{l, AL} &={} -\frac{\bar{g}^2l^2 q_0^{\frac13}}{16 \pi^4 v_F^4 \gamma^{\frac23}}\int d\abs{\bm{p}}\, \abs{\bm{p}}^3\left[(\abs{\bm{p}}^3+1) \ln \left(1+\frac{1}{\abs{\bm{p}}^3}\right)-1 -\frac{1}{2\abs{\bm{p}}^3}\right] = +\frac{\bar{g}^{\frac43} l^2 q_0^{\frac13}\sqrt{3}}{7{\pi}^{\frac{7}{3}}2^{\frac{13}{3}} v_F^{\frac{10}{3}}m^{\frac23}}\,,
    \end{split}
  \end{equation}
\end{widetext}
which again exactly cancels with the MT+SE contribution of Eq.~\eqref{Eq:pimtse_qcp2}.

\subsubsection{Subleading contribution from AL diagrams in odd channels}
In the previous sections, we have shown that the leading contributions from the SE+MT and AL diagrams cancel each other, and that there are no subleading contributions from the SE+MT diagrams besides the regular ones ($\xi$-independent) which are irrelevant for our purpose. For this reason, the polarization bubble in  odd angular momentum channels is determined by subleading terms originating from the AL diagrams.

To streamline the calculation, we introduce a dimensionless parameter:
$w=\xi^{-2}(q_0/2)^{-\frac23}\gamma^{-\frac23}$, such that the QCP corresponds to $w=0$, and
the FL regime to $w \to \infty$, rescale the internal bosonic frequency and momenta as $\abs{\bm{p}} \to \abs{\bm{p}} (q_0/2)^{\frac13} \gamma^{\frac13}$ and $p_0 \to \zeta q_0/2$, and  denote $\eta_{\pm}=\abs{\zeta \pm 1}$. The polarization bubble for odd $l$ then takes the form:
\begin{equation}
\label{Eq:pistab}
    \Pi_l (q_0) = -\bar{g}^2 \frac{l^4 q_0^{5/3}}{v_F^6 \pi^4 2^{\frac{26}{3}} \gamma^{\frac43}} F(w)\,,
\end{equation}
where
\begin{widetext}
    \begin{equation}
    \label{Eq:theF}
        \begin{split}
            F(w)&={}\int_0^\infty \frac{d\abs{\bm{p}}}{\abs{\bm{p}}} \int_0^\infty \! d\zeta \frac{1}{\bm{p}^2+w+\frac{\eta_+}{\abs{\bm{p}}}}\frac{1}{\bm{p}^2+w+\frac{\eta_-}{\abs{\bm{p}}}}{({\eta_{+}}-{\eta_{-}})}^2 (2\eta_{+}^2+3{\eta_{+}}{\eta_{-}}+2\eta_{-}^2)=\\
            &={} \int_0^\infty\!\! \abs{\bm{p}} d\abs{\bm{p}} \int_0^1 \! d\zeta \left(\frac{4\zeta^2(\zeta^2+7)}{(\abs{\bm{p}}^3+w \abs{\bm{p}}+\zeta+1)(\abs{\bm{p}}^3+w \abs{\bm{p}}+1-\zeta)}-\frac{28\zeta^2}{{(\abs{\bm{p}}^3+w \abs{\bm{p}}+\zeta)}^2}\right)+\\
            &+{}\int_0^\infty\!\! \abs{\bm{p}} d\abs{\bm{p}} \int_1^{\infty} \! d\zeta \left(\frac{28\zeta^2+4}{(\abs{\bm{p}}^3+w \abs{\bm{p}}+\zeta+1)(\abs{\bm{p}}^3+w \abs{\bm{p}}+\zeta-1)}-\frac{28\zeta^2}{{(\abs{\bm{p}}^3+w \abs{\bm{p}}+\zeta)}^2}\right)\,.
        \end{split}
    \end{equation}
\end{widetext}

We once again added a regular counter-term $28\zeta^2/(\abs{\bm{p}}^3+w \abs{\bm{p}}+\zeta)^2$ to regularize the frequency integral. It is easy to see that this counter-term contributes only to a static, frequency-independent part of the bubble. Indeed, rescaling $\abs{\bm{p}} \to \sqrt{w}\abs{\bm{p}}$ and $\zeta \to w^{\frac32} \zeta$, we find the contribution from this term to $F(w)$ in the form
\begin{equation}
  \begin{split}
    &\int_0^\infty\!\! \abs{\bm{p}} d\abs{\bm{p}} \int_0^{\frac{\mathcal{L}}{q_0}} \! \frac{\zeta^2 d\zeta}{{(\abs{\bm{p}}^3+ w \abs{\bm{p}}+\zeta)}^2}=\\
    &=w^{\frac52} \int_0^\infty\!\! \abs{\bm{p}} d\abs{\bm{p}}  \int_0^{\mathcal{L}\xi^3} \! \frac{\zeta^2 d\zeta}{{(\abs{\bm{p}}^3+\abs{\bm{p}}+\zeta)}^2}\,,
  \end{split}
\end{equation}
where $\mathcal{L}$ is the UV cutoff. Substituting into~\eqref{Eq:pistab} and using $w^{\frac52} \sim q_0^{-\frac53}$, we find that this term contributes only a constant to $\Pi_l (q_0)$ and does not affect its frequency dependence.

At large $w$ (the FL regime), the scaling function behaves as:
\begin{equation}
      F(w)  \sim \frac{\frac{20 \pi}{3 \sqrt{w}}}{1+\frac{16}{25 \pi w^{\frac32}}\ln w^{\frac32}} \approx \frac{20 \pi}{3 \sqrt{w}} -\frac{64}{15 w^2}\ln w^{\frac32}\,.
\end{equation}
Substituting the last expression into Eq.~\eqref{Eq:pistab}, we obtain the full contribution to the polarization for odd $l$ in the FL regime:
\begin{equation}
\label{Eq:AL_FL_full}
  \begin{split}
    \Pi_{l} &= {}-\frac{5\bar{g}}{192 \pi^2 m v_F^5}\frac{l^4 q_0^2 \xi}{1 +  \frac{4\bar{g} m}{25 \pi^2 v_F}q_0 \xi^3 \ln{\frac{1}{\xi^3 q_0}}}\\
   &\approx {}-\frac{5\bar{g}}{192 \pi^2 m v_F^5}l^4 q_0^2 \xi+   \frac{\bar{g}^2}{240 \pi^4 v_F^6}l^4 q_0^3 \xi^4 \ln \frac{1}{\xi^3 q_0}\,.
  \end{split}
\end{equation}

Rotating  now to real frequencies, we obtain
\begin{equation}
\label{Eq:AL_FL_im}
   \Pi''_{l} (\omega) = + \omega^3 \bar{g}^2 \frac{l^4 \xi^4 }{240 \pi^4 v_F^6} \ln \frac{1}{\xi^3 \omega}\,.
\end{equation}
The positive sign of $\Pi''_{l, AL}$ implies that the mode dissipates energy, rather than grows in time, indicating that the system is stable.

\begin{figure}
  \includegraphics[width=0.9\linewidth]{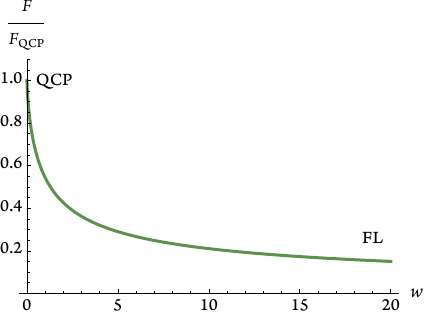}
  \caption{The scaling function, normalized to its value at QCP\@. \label{Fig:matsubara}}
\end{figure}

At the QCP ($u=0$), the integrals in Eq.~\eqref{Eq:theF} yield:
\begin{equation}
  \Pi_{l} (q_0) ={} -\frac{49 \sqrt{3}}{3520\pi^3}\frac{\bar{g}^2l^4 q_0^{\frac53} }{ v_F^6 \gamma^{\frac43}}\,.
\end{equation}

Rotating to real frequency, we obtain
\begin{equation}
\label{Eq:AL_QCP_im}
    \Pi''_{l}(\omega) ={} +\sin{(\tfrac{5\pi}{6})}\frac{49 \sqrt{3}}{3520\pi^3}\frac{\bar{g}^2l^4 }{ v_F^6 \gamma^{\frac43}} \abs{\omega}^{\frac53} \sgn{\omega}\,,
\end{equation}
We see that $\sgn{\Pi''_{l}(\omega)} = \sgn{\omega}$, as requires for causality. Analyzing the susceptibility $\chi_l (z)$ in the upper half-plane of frequency,  $z= \omega' + i \omega^{''} = |z| e^{i\phi}$, $0<\phi <\pi$, we find
\begin{equation}
\label{Eq:causal_susceptibility}
  \chi_l(z) \propto \frac{1}{1 + a_l \frac{49 \sqrt{3}}{3520\pi^3}\frac{\bar{g}^2l^4 }{ v_F^6 \gamma^{\frac43}} \abs{z}^{\frac53}e^{i\frac53 (\phi -\frac{\pi}{2})}}\,,
\end{equation}
This $\chi_l (z)$ has no poles in the upper half-plane. This guarantees the causality of collective excitations.

The full scaling function $F(w)$ for arbitrary $w$ is plotted in Fig.~\ref{Fig:matsubara}. It remains sign-definite across its domain, indicating that the polarization bubble in Eq.~\eqref{Eq:pistab} does not change sign for any finite $\xi$, reinforcing the statement about the stability of the system at any distance from the QCP\@.

The full complex $\Pi_l (\omega)$  on the real axis  at arbitrary $\xi$ is
\begin{equation}
\label{Eq:pistab_re}
  \Pi_l (\omega) = \bar{g}^2 \frac{l^4 \omega^{5/3}}{v_F^6 \pi^4 2^{\frac{26}{3}} \gamma^{\frac43}} {\bar F} (x)\,,
\end{equation}
where $x=\xi^{-2}(\omega/2)^{-\frac23}\gamma^{-\frac23}$, and ${\bar F}(x)=e^{i\frac{\pi}{6}} F(e^{i\frac{\pi}{3}} x)$. In Figure~2 in the main text we  show the imaginary part of ${\bar F} (x)$, normalized to its value at the QCP\@. The function remains sign-definite for all $x$, confirming that dissipation does not reverse sign anywhere along the crossover from the FL to the QCP\@.

\subsection{Additional remarks about the renormalization of the side vertices}
\label{app_B}

In the main text we argued that to leading order in $1/k_F\xi$, the renormalization factor $\Gamma_l$ for a side vertex in a dynamical uniform particle-hole polarization bubble scales as $1/Z = 1 + \lambda$. Analyzing vertex renormalization more carefully, we find that this holds for the terms linear in $s_{\pm}$ in $\mathcal{P}(p,q)$ in Eq.~\eqref{Eq:ferm_expansion} as for such terms, $\sin^2{\theta}$ coming from the side vertex, can be approximated by its value at $\theta = \pi/2$ (see Eq.~\eqref{Eq:mt_vertex_expansion}). For the terms of higher-order in $s_{\pm}$ in $\mathcal{P}(p,q)$ for the AL diagrams (the ones which account for the leading contribution to $\Pi_l$ for odd $l$) the situation is different as these terms come from the expansion of  $\sin\theta_{\bm{k}\bm{p}} \sin{l\theta_{\bm{k}\bm{p}}}$ around $\pi/2$, i.e., around $\bm{k}$ perpendicular to $\bm{p}$. Let's set $l=1$ for definiteness.  The subleading term is then $k^2_{\parallel}$. Typical $k_{\parallel} \sim  \abs{\omega}/(v_F\xi^{-1})$. Meanwhile, typical bosonic momenta in each  term in ladder series of vertex renormalizations are of order $1/(k_F \xi)$. Under the condition $\omega \ll \xi^{-2}/m$, which is well satisfied in a FL regime~\footnote{A FL behavior holds when $\gamma \omega \ll \xi^{-3}$. This is equivalent to $\omega \ll (\xi^{-2}/m) /\lambda$. Combining this with  $\lambda \gg 1$, we find that  the condition $\omega \ll \xi^{-2}/m$ is well satisfied.} the subleading  term in the form-factor cannot be taken out of the integral over bosonic momenta in the  vertex correction diagram in Fig.~3 of the main text. An order-of-magnitude analysis shows that in this situation the integral over bosonic propagator  does not compensate the small  overall factor of $Z$ coming from the fermions. As the consequence, the vertex correction is small and the fully dressed prefactor for $k^2_{\parallel}$ nearly coincides with the bare one.\\

\subsection*{Renormalization of $\Pi_l (\omega)$  with odd $l$ in the FL regime}

As we just found, for odd $l$, there is no singular vertex renormalization, but there are two extra powers of $s_{\pm} \propto 1/Z$ in $\mathcal{P}(p,q)$ in Eq.~\eqref{Eq:ferm_expansion}. As a result, we again have
\begin{equation}
  \Pi_{l} (\omega) = \Pi^{Z=1}_{l} (\omega) \frac{Z^2}{Z^2} = \Pi^{Z=1}_{l}
\end{equation}

\bibliography{paper}

\end{document}